\documentclass[twocolumn]{aastex63}

\usepackage{amssymb}
\usepackage{amsmath}

\shorttitle{Disintegrating Exoplanets}
\shortauthors{Baka et al.}

\graphicspath{{./}{Lightcurves/}}

\begin{document}

\title{Disintegrating Exoplanets: Creating Size Constraints by Statistically Peering Through the Debris
\footnote{Released on \today}}
 
\correspondingauthor{Keith M Baka}
\email{keithbaka17@gmail.com}

\author{Keith M Baka}
\affiliation{Steward Observatory, The University of Arizona, 933 North Cherry Avenue, Tucson, AZ 85721, USA
}
\affiliation{Anton Pannekoek Institute for Astronomy, The University of Amsterdam, Science Park 904, 1098 XH, Amsterdam, The Netherlands}

\author{Everett Schlawin}
\affiliation{Steward Observatory, The University of Arizona, 933 North Cherry Avenue, Tucson, AZ 85721, USA
}

\begin{abstract}

We study two intriguing exoplanets, Kepler-1520b and K2-22b, that are being disintegrated by their host stars, producing dust and debris pulled from their surface into tails that trail and precede the exoplanets in their orbits, making it difficult to discern the true nature of the objects. Our goal is to constrain the radius of the underlying objects, and while previous studies have done this in the past by selecting shallow transit events, we attempt a new statistical approach to model the intrinsic astrophysical and photon noise distributions simultaneously. We assume that the lightcurve flux distribution is distributed as a convolution of a Gaussian photon noise component and a Raleigh astrophysical component. The Raleigh curve has a finite flux maximum, which we fit with a Hamiltonian Markov Chain. With these methods, a more accurate flux maximum may be estimated, producing a better final value for the radius of these exoplanets. To determine statistical significance, we used the python package PyMC3 to find the posterior distribution for our data with Gaussian, Rayleigh, and joint function curves and plotting it against our collected flux.


After completing this analysis, we were able to place a constraint on the maximum radius of K2-22b at approximately $15,000km$. However, we were unable to constrain the radius of Kepler-1520b due to either transits obscured by forward scattering of dust or a grazing transit.

\end{abstract}


\section{Introduction} \label{sec:intro}

Kepler-1520 is a star located approximately 620pc from Earth with an exoplanet candidate with a period of 0.65 days orbiting it. The Kepler observatory collected long cadence photometry of Kepler-1520 for 4 years (\citet{borucki2010}). In 2012, a study conducted by \citet{rappaport2012} revealed that the orbiting exoplanet was a candidate for being a disintegrating exoplanet, as the light curve has flux dips ranging between $\sim$0\% and 1.3\%, while maintaining a highly regular interval of approximately 0.654 days (\citet{Budaj13}). The coincidence of the flux variability and the asymmetric shape of the transit makes it likely that Kepler-1520b is a disintegrating rocky exoplanet with a variable comet-like tail of debris trailing behind it.

Another similar exoplanet, K2-22b, was first investigated in 2015 by \citet{sanchis2015}. This study showed an exoplanet orbiting its host star with a consistent period of 0.381 days while producing transit depths between 0\% and 1.3\%. Additionally, analysis of the light curves revealed features at the ingress and egress points for the exoplanet transit consistent with both a leading and following dust trail. With these two traits, it is likely that K2-22b is another a disintegrating rocky exoplanet similar to Kepler-1520b.

Previous studies by \citet{perez13} and \citet{booth23} have found that small, rocky planets that have entered into this mass loss regime are experiencing catastrophic mass loss that creates highly variable clouds of dust. The 2013 work by \citet{perez13} indicates that Kepler-1520b, and by our assumptions, K2-22b, are in the last few Gyr of their lifetimes, having been stripped of a majority of their mass already. From work by \citet{booth23}, the variability of these dust clouds is likely caused by the evaporation and condensation timescales of dust grains, which are dependent on the surface temperature of the planet and the resulting outflow.

The 2018 review by \citet{vanL2018} estimated the minimum size of both planets. By analyzing the variability in the extinction due to the debris trail, they were able to determine the mass-loss rates for planets of different sizes that produce a debris trail of that size. Through this and the assumed lifetime of the planet of under 500 Myr, \citet{vanL2018} estimated that both planets are of size on the order of 500km.

\begin{figure}[t]
\centering
\includegraphics[width=\columnwidth]{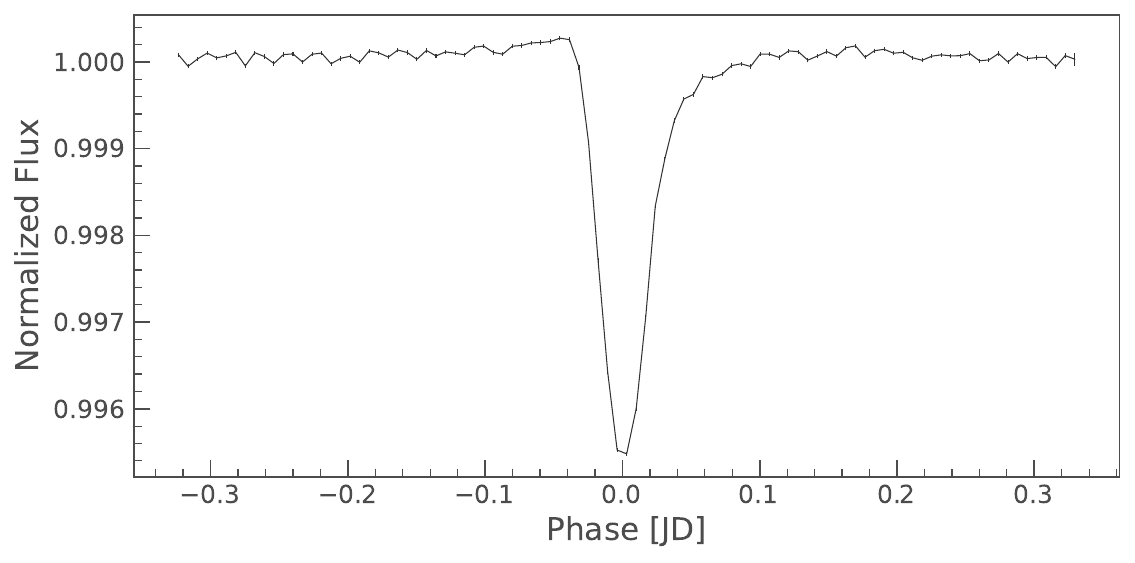}
\caption{Phase-folded and binned lightcurve for the transit of Kepler-1520b. This data, collected by Kepler, contains approximately 1773 transits of the exoplanet. This lightcurve is the average of all those transits, and clearly shows a transit profile that is unusual for an exoplanet, as it is asymmetric and shows an increase in flux preceding transit. Standard error of the mean is shown for each bin. \label{kepfig1}}
\end{figure}

While previous papers, such as \citet{rappaport2012} and \citet{sanchis2015}, have studied the approximate size and nature of the exoplanets in conjunction with their trail(s) of debris, the goal of this paper is to look through the veil of the debris trails in order to characterize the underlying planets to place a constraint on the radius of each. Another technique used by other papers is to study quiescent transits, or transits with low dust activity \citep{sanchis2015}, \citep{rappaport2012} in order to place a minimum value on the radius of the planets. This approach often gives good results, but can be subject to error in that the selected transits may not be the minimum dust activity, or may only appear to be shallow transits due to random statistical photon scatter. The radius of these planets can help derive other properties, such as the dynamics and the escape mechanism.

The goal of this study is to put a tighter constraint on the maximum size of both planets. We will attempt to unpack the clouds of debris that surround the planets with precision in order to obtain a better understanding of the systems in their entirety. In section 2, we discuss how we modeled the transits of the disintegrating exoplanets, and in section 3 we discuss the results of our analysis.



\section{Data Reduction} \label{sec:data}

\subsection{Observations} \label{lightcurveD}

We use the publicly available data from the Kepler mission, with Kepler release 25 for Kepler-1520b and K2 data release 32 for K2-22b, and download the long cadence, SAP flux for both planets, as the short cadence data did not cover as many transits as the long cadence data and thus had lower statistical power. To do this, we used the program \texttt{lightkurve} \citep{Lightkurve} to download the data and produce a FITS file of the transits.

\begin{figure}[t]
\includegraphics[width=\columnwidth]{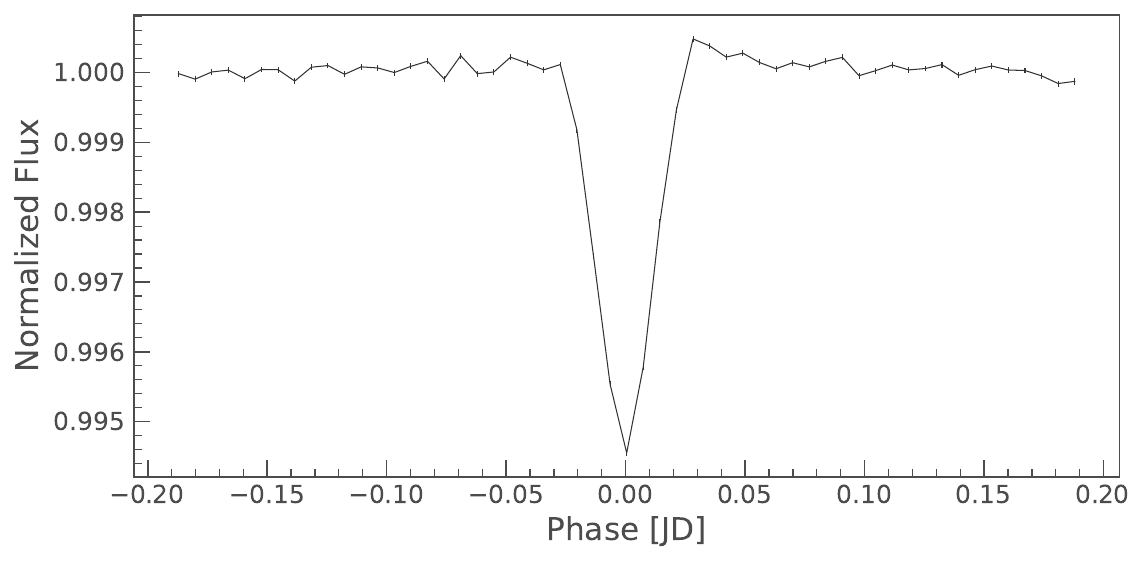}
\caption{Phase-folded and binned lightcurve for the transit of K2-22b. This data, collected by Kepler, contains 202 transits of the exoplanet. This lightcurve is the average of all those transits, and while it does not clearly show an asymmetry during the primary transit, the bump in flux above 1.0 following the egress is indicative of forward scattering by dust. Standard error of the mean is shown for each bin.
\label{k2fig1}}
\end{figure}

The next piece of code produced phase-folded, cleaned lightcurves via several steps. First, we normalized the flux of the data and removed any outliers that were above five times the standard deviation to remove cosmic rays, and we do this for the entire time Kepler observed the systems. We also used the lightkurve flatten function with a window length parameter of 101 to remove stellar activity in this step and produce flattened lightcurves. We then used a box least squared periodogram to find the best period for each planet, using results from \citet{rappaport2012} and \citet{sanchis2015} to place prior constraints on our periodogram, selecting the primary peak of the periodogram as the period for each planet, giving a period of 0.6536 days for Kepler-1520b and 0.3811 days for K2-22b. The next step was to use the period we obtained for each planet to phase-fold the lightcurve, and then finally bin the result to produce the results seen in Figure \ref{kepfig1} and Figure \ref{k2fig1}. The asymmetry of the lightcurves agrees with results from \citet{rappaport2012} and \citet{sanchis2015}.

\subsection{Time Slicing and Folding Data} \label{subsec:Slices}

%

\begin{figure*}[b!]
    \centering
    \includegraphics[width=\columnwidth]{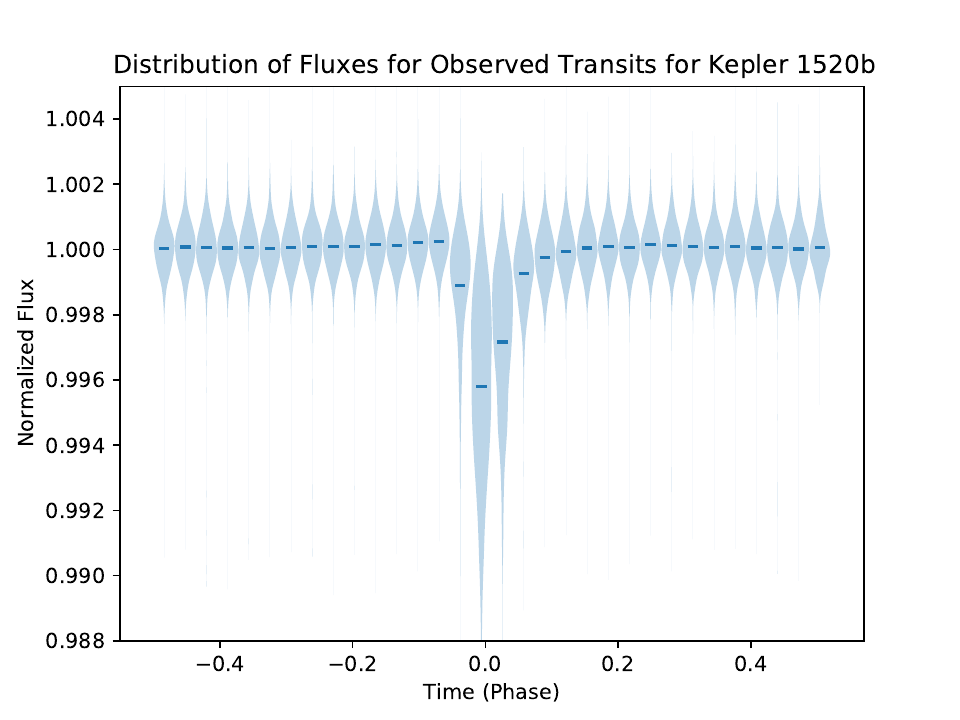}
    \centering
    \includegraphics[width=\columnwidth]{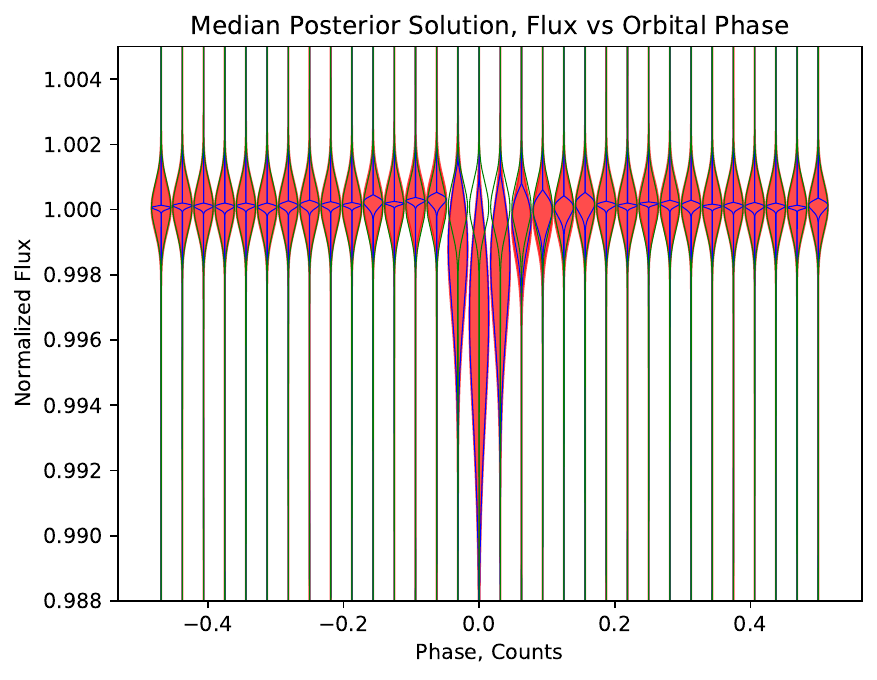}
    \centering
    \includegraphics[width=\columnwidth]{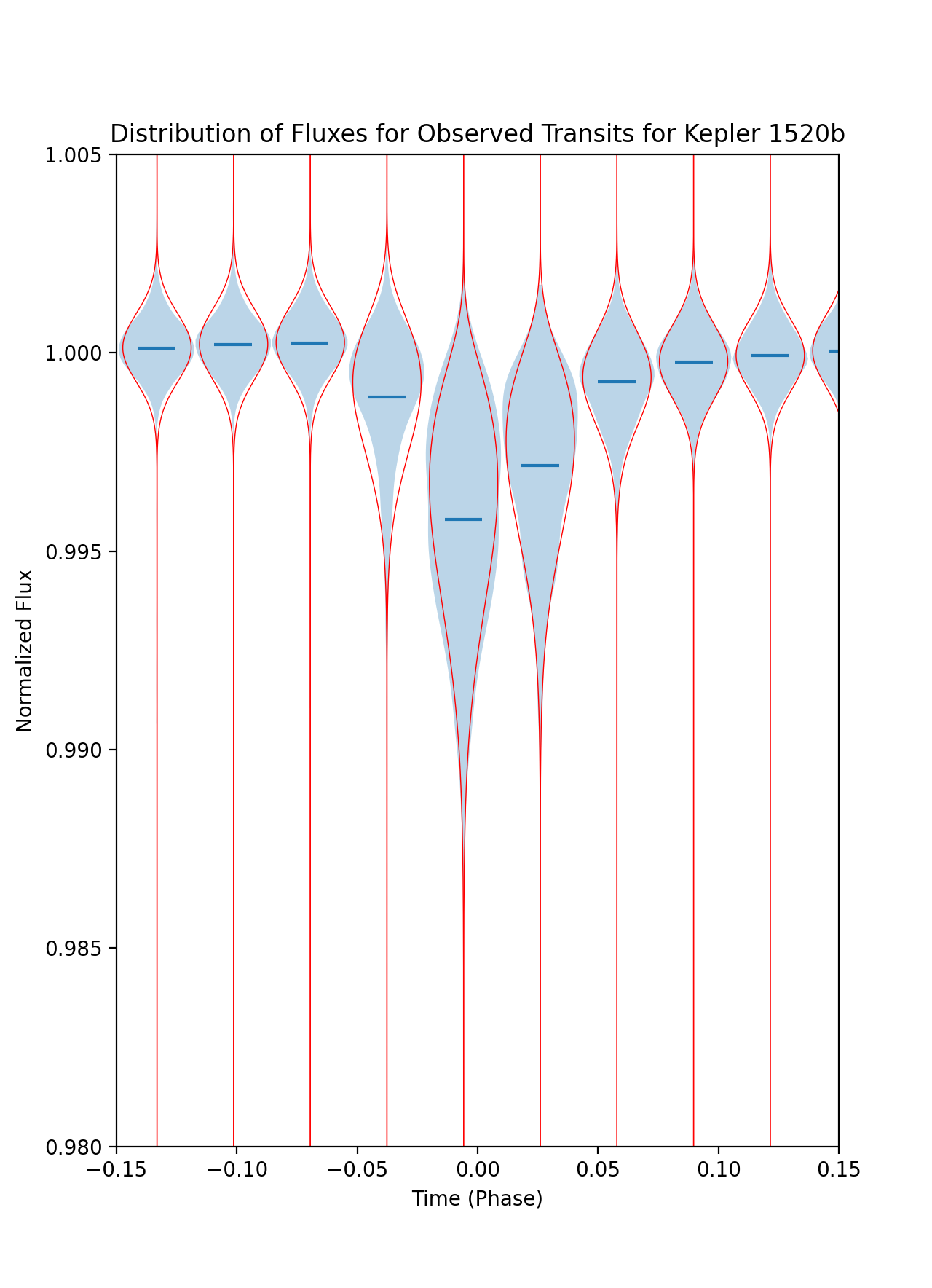}
    \caption{Violin plot showing the observed flux for the orbit of Kepler-1520b, where each violin represents all the fluxes from a 30-minute segment of the orbit, and the blue bars signify the mean of each segment. The violins near the center show the transit and slow egress of the planet and its dust trail (Left). Violin plot of the median of the posterior solution for Kepler-1520b, where the salmon filled section indicates the solution for the joint function, green lines indicate the Gaussian component of the solution, and the blue lines indicate the Rayleigh component of the solution (Right). Comparison between data (blue) and maximum a priori model (red) for the in-transit data of Kepler-1520b (Bottom).}
    \label{kepviolin1}
\end{figure*}

\begin{figure*}[t!]
    \centering
    \includegraphics[width=\columnwidth]{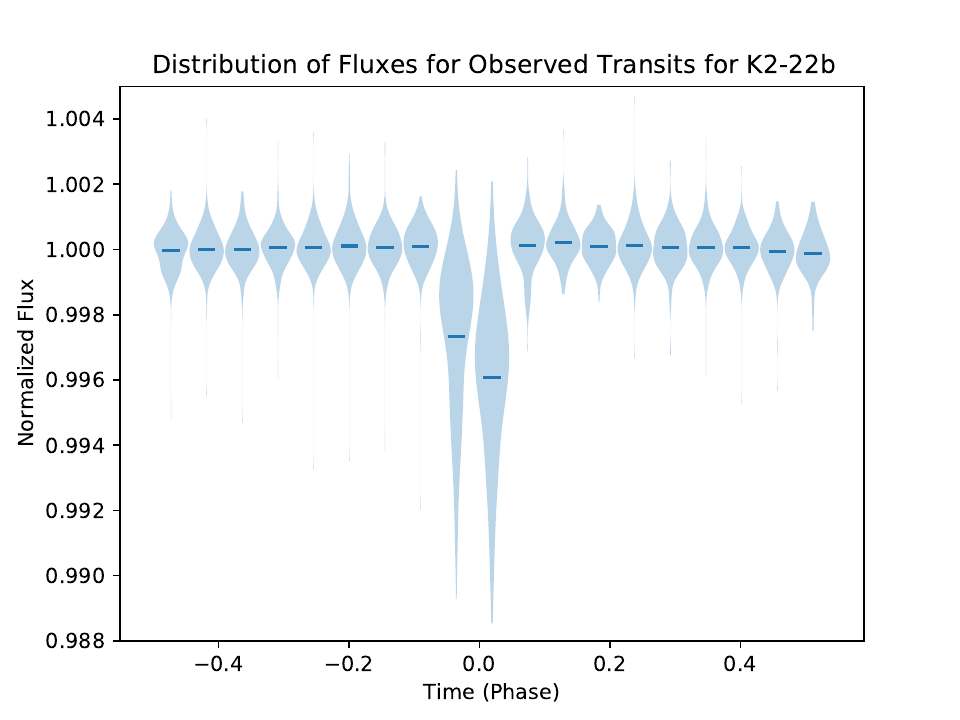}
    \centering
    \includegraphics[width=\columnwidth]{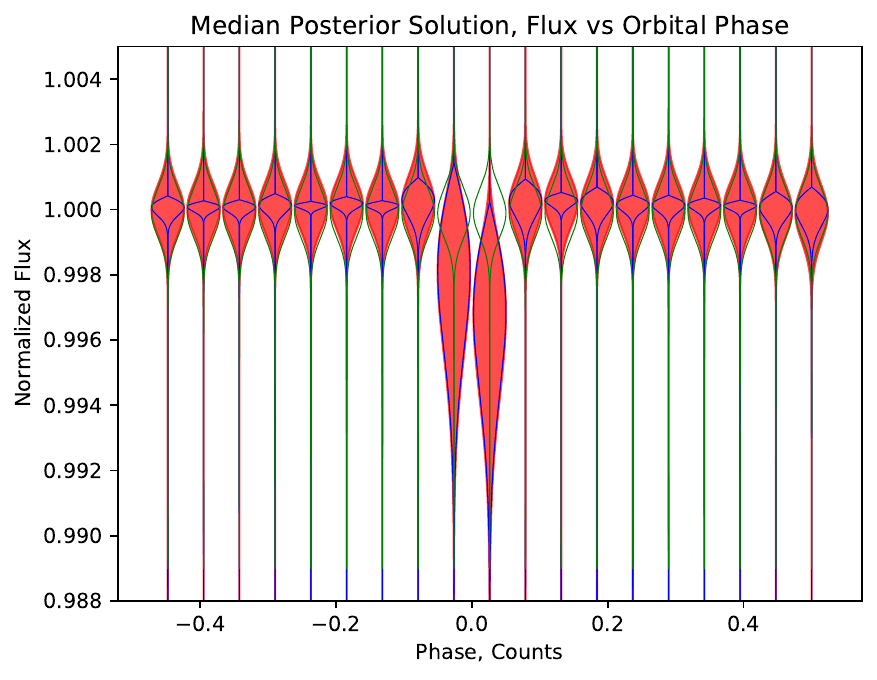}
    \centering
    \includegraphics[width=\columnwidth]{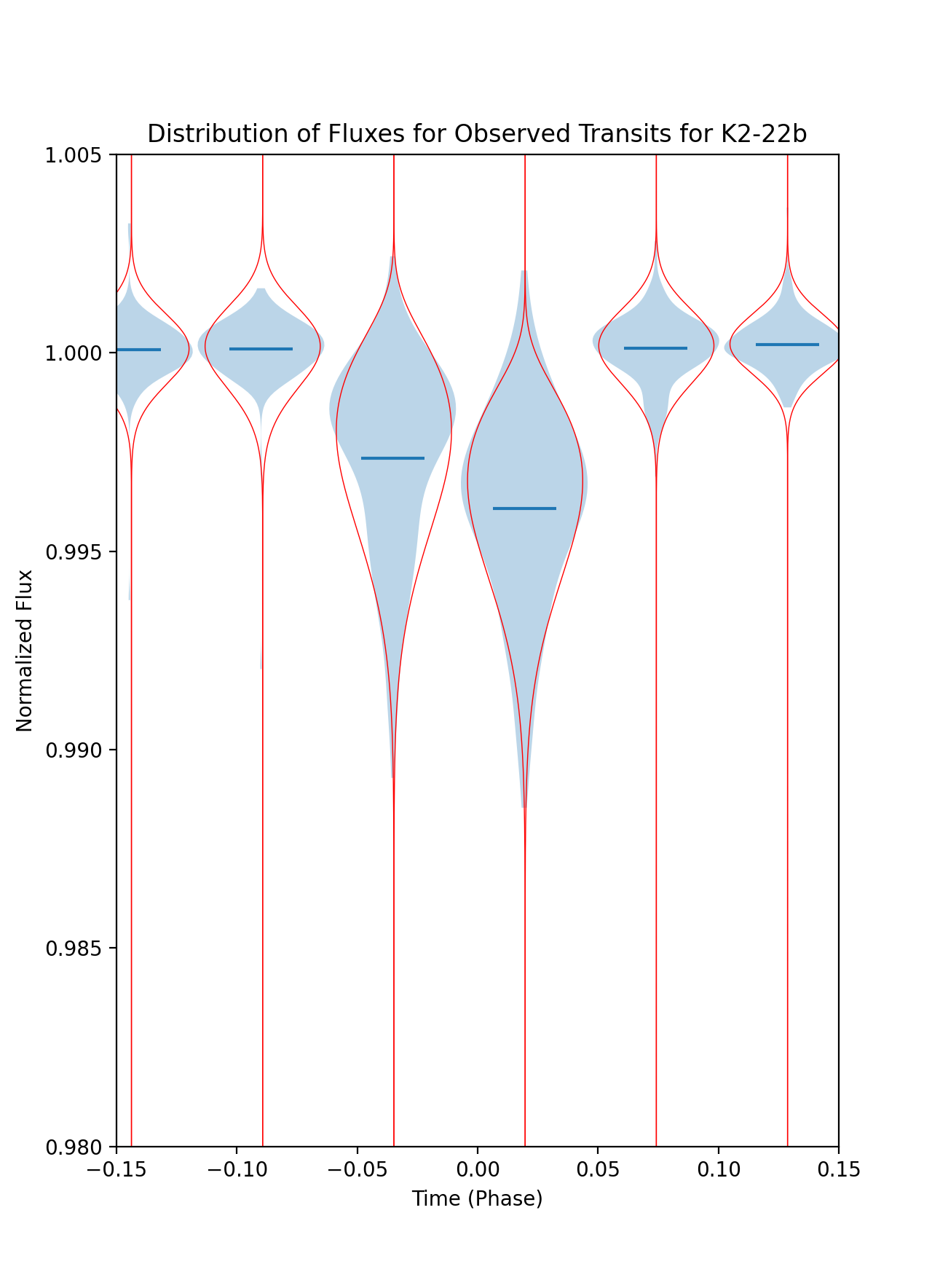}
    \caption{Violin plot showing the observed flux for the orbit of K2-22b, where each violin represents all the fluxes from a 30-minute segment of the orbit, and the blue bars signify the mean of each segment. The violins near the center show the transit and slow ingress of the planet and its leading dust tail (Left). Violin plot of the median of the posterior solution for K2-22b, where the salmon filled section indicates the solution for the joint function, green lines indicate the Gaussian component of the solution, and the blue lines indicate the Rayleigh component of the solution (Right).Comparison between data (blue) and maximum a priori model (red) for the in-transit data of K2-22b (Bottom).}
    \label{k2violin1}
\end{figure*}

With the knowledge that the Kepler probe recorded data once every 30 minutes, we then used the periods we obtained to split the binned data into 30 minute slices, 32 for Kepler-1520b and 19 for K2-22b. We wrote a code that saved the flux for each time slice, binned by flux with a bin size of 0.0005 (normalized flux).
This information is then written to a FITS file, and plotted in violin plots which can be seen on the left side of Figures \ref{kepviolin1} and \ref{k2violin1}. We also repeated this process for 15 minute time slices, which Nyquist sample the 30 minute time slices, but did not plot this data as it was to narrow to view in this format.


\subsection{Joint Function for Photon and Astrophysical Noise \label{joint_func}}

Our probability distribution function is a convolution of a Gaussian, to account for photon noise, and a Rayleigh, to model the astrophysical variation. We will assume that \begin{enumerate}
    \item total extinction is greater than total forward scattering,
    \item dust in the system occults the star
    \item astrophysical variations follow a Rayleigh distribution,
\end{enumerate}

we consider the photon uncertainty to follow a Gaussian probability distribution function (PDF):
\begin{equation}\label{eq:Gaussian}
f_G(x) = \frac{1}{\sqrt{2 \pi} \sigma_G} e^{\frac{-(x-\mu_G)^2}{2 \sigma_G^2}},
\end{equation}
where $\mu_G$ is the mean of the distribution and $\sigma_G$ is the standard deviation.

We assume that the dust tail can only decrease the flux mid-transit and therefore follows a PDF with a sharp cutoff, following assumption 1. We assume that it follows a Rayleigh distribution:
\begin{equation}
f_R(x) =
\begin{cases}
	 \frac{\mu_R - x}{\sigma_R^2} e^{\frac{-(\mu_R - x)^2}{2 \sigma_R^2}} & x < \mu_R \\
	 0 & x \ge \mu_R,
\end{cases}
\end{equation}
where $\sigma_R$ is the Rayleigh scale parameter and $\mu_R$ is the maximum of the flux distribution, which is near 1.0.
The PDF of the sum of a random variable following the Rayleigh and Gaussian PDFs will be the convolution.
\begin{equation}\label{eq3}
(f * g) (x) = \int_{-\infty}^{\infty}{f_G(x - u) f_R(u) du}
\end{equation}

\begin{equation}\label{eq4}
(f * g) (x) = \frac{f_G(x\prime)}{1 + \frac{\sigma_R^2}{\sigma_G^2}} + \frac{\sigma_R}{2 \sigma_2^3} \left (x\prime  e^{\frac{-x\prime^2}{2 \sigma_2^2}} erfc\left( \frac{- x\prime \sigma_R}{\sqrt{2} \sigma_2 \sigma_G} \right) \right),
\end{equation}
where
$x\prime = x - \mu_R$,
\begin{equation}
\sigma_2 = \sqrt{\sigma_R^2 + \sigma_G^2}
\end{equation}
and $erfc$ is the complementary error function:
\begin{equation}
erfc(x) \equiv 1 - \frac{2}{\sqrt{\pi}} \int_{0}^{x} e^{-x^2} dx,
\end{equation}
and x is the random variable for the flux.


\subsection{Bayesian Analysis}

To calculate the hyperparameters of this distribution (and thus the maximum flux and minimum dust extinction), we used the software package \texttt{pymc3} \citep{pymc3} so that we could investigate the maximum a priori solutions as well as the posterior distributions of the models we designed for both planets.

We started by modeling the out-of-transit data, which included observations taken before phase = -0.2 and after phase = 0.2, using a Gaussian prior as per equation \ref{eq:Gaussian} for both the mean and standard deviation. As a result of the normalization process we used (all lightcurves were normalized by the mean out-of-transit), we created a prior on $\mu_G$ held at the same value at $\mu_{G,P} = 1$, where $\mu_{G,P}$ is the prior on the out-of-transit mean, and a prior on the standard deviation equal to the photon noise of the out-of-transit data.

For the 30 minute time bins described in section \ref{subsec:Slices}, we returned to the joint Gaussian and Rayleigh function we derived in \ref{joint_func}, using equation \ref{eq4}. In order to create a model prior using this function, we created PyMC3 random variable objects for two variables as seen in equation \ref{eq4}: the mean and standard deviation of the Rayleigh distribution, $\mu_R$ and $\sigma_R$. For the Rayleigh distribution, we chose the prior on the mean to be $\mu_R = 1$ and the Rayleigh scale parameter to be $\sigma_R = 0.01$. For the Gaussian component of the in-transit data, we used the results for the out-of-transit Gaussian data, holding $\mu_G$ constant and only allowing the Rayleigh to shift and broaden. 

The joint Raleigh-Gaussian functions for the median parameters for $\mu_R$ and $\sigma_R$ for each planet can be found in the right hand side of Figures \ref{kepviolin1} and \ref{k2violin1}. In these figures, the fixed Gaussian can be seen in green, the variable Rayleigh function can be seen in blue, and the joint function is filled in with salmon.

\section{Results \& Discussion} \label{sec:results}

\subsection{Flux Maxima and Planet Radius Constraints}

For our results, we are looking at the posterior distribution of the variable $\mu_R$, which is the maximum of the PDF. This is representative of the underlying lightcurve, where the variable aspect of the dust has been removed.

\subsubsection{Kepler-1520b} \label{subsec:DatKep}

\begin{figure}[t!]
\centering
\includegraphics[width=\columnwidth]{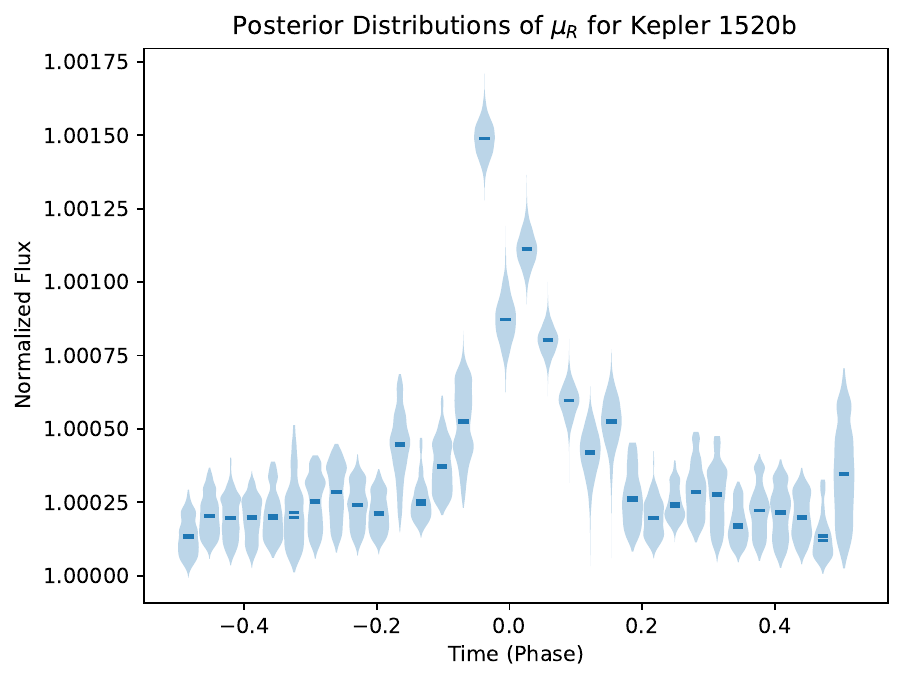}
\caption{Violin plots of $\mu_R$ for the posterior distributions of the joint function for Kepler-1520b, where $\mu_R$ is the maximum of the Rayleigh component of the flux distribution. Each violin represents all the $\mu_R$ from a 30 minute segment of the orbit.
\label{kepfig3}}
\end{figure}


The results we obtained for Kepler-1520b leave lots of room for interpretation and potential further study. First, when examining the posterior distributions, several interesting patterns emerge. Starting with Figure \ref{kepfig3}, we see that during the transit, the maximum value for the flux actually increases. Thus, assumption 1 in section \ref{joint_func} appears to be invalid and forward scattering can be the dominant effect.
Right at $Phase = 0$, we can see there is a slight dip before the flux increases again. This may be due to the transit of Kepler-1520b, but it is probably still obscured partially by forward scattering. In case these time slices were too wide, and the values for maximum flux were a result of starlight scattering from the dust trail either before or after the transit, we chose to make another version of this plot with smaller time slices.

\begin{figure}[t]
\centering
\includegraphics[width=\columnwidth]{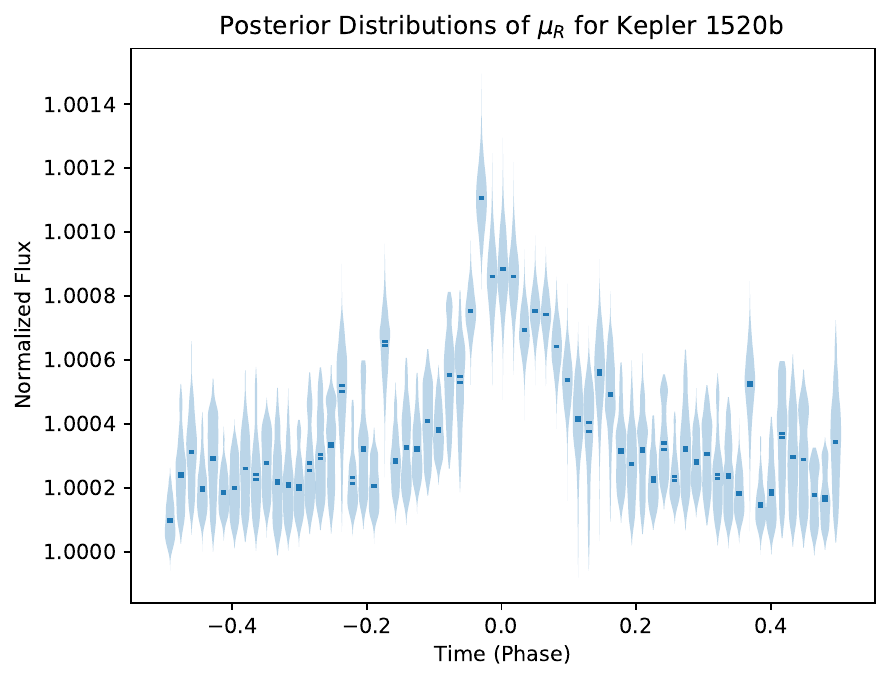}
\caption{Violin plot of $\mu_R$ for the posterior distributions of the joint function for Kepler-1520b, where $\mu_R$ is the maximum of the Rayleigh component of the flux distribution. Each violin represents all the $\mu_R$ from a 15 minute segment of the orbit.
\label{kepfig2}}
\end{figure}

We can also examine smaller time slices of half the exposure time, which allows us to study the time behavior at the Nyquist rate. After remaking the posterior distribution with 15 minute time slices, shown in Figure \ref{kepfig2}, we see that the dip in flux that was visible in \ref{kepfig3} is no longer visible. Additionally, there is a slower build to the transit ingress, suggestive of a small leading debris trail scattering light to the detectors.

To better understand the whether the dip seen in Figure 5 could be astrophysically related to the planet, we compare the 15 and 30 minute segment sizes to the expected duration of the transit. We can use the equation
\begin{equation}\label{eq5}
    T_{tot} = \frac{P}{\pi} sin^{-1}[\frac{R_{*}}{a} \frac{\sqrt{(1+k)^2-b^2}}{sin(i)}],
\end{equation}
where
\begin{equation}
    k = \frac{R_p}{R_*},\; b = \frac{acos(i)}{R_*}
\end{equation}
from \citet{seagerbook2010}, and use stellar parameters from \citet{rappaport2012} and the exoplanet radius value from \citet{vanL2018} to obtain values for the transit durations, assuming a circular orbit. Doing this, we find that with impact parameters of 0, 0.5, and 1, the transit times are 70.4, 61.2, and 6.8 minutes respectively, or 0.075, 0.065, and 0.0036 of the orbit, meaning that for all but a nearly grazing transit, our slices should effectively time resolve the transit of Kepler-1520b.

In both versions of the time series though, one important pattern is visible. We can see in both graphs (Figures \ref{kepfig3} and \ref{kepfig2}) that the $\mu_R$ variable is elevated above the out-of-transit value well after the egress of the planet that culminates in a plateau before returning to normal, suggesting a large debris tail that maintains a significant size even well after the planet has transited. 

Assuming that the dip present in Figure \ref{kepfig3} is a result of a true planet transit causing extinction amongst the forward-scattered light, we can say that the planet blocked an average of 0.043\% of the flux coming from the star. Thus, using the equation
\begin{equation}
\label{eq1}
    \Delta F = (R_{Planet}/R_{Star})^2
\end{equation} along with $ R_{Star} = 0.65R_\odot\ $ (\citet{rappaport2012}), we can calculate that the maximum possible radius for Kepler-1520b is approximately 9400 km, which is much larger than the upper limit reported by \citet{vanW} using an assumed albedo and the upper limit of the eclipse, while our method is independent and does not make assumptions about the albedo of the planet. 

However, this constraint on the radius is contingent on there being a large cloud of dust forward scattering light, that the transit of the planet is actually the cause of the dip in the value of $\mu_R$, and that our model is correct. Since the value of $\mu_R$ is elevated above unity for the duration of the transit, this means that the first or second assumptions for our model are incorrect. Additionally, the dip seen in Figure \ref{kepfig3} is not found in Figure \ref{kepfig2}, meaning that the result is not robust. Furthermore, were this dip truly caused by the planet, the maximum radius of 9600km is much larger than the upper limit reported by \citet{vanW}, leaving the possibility that there is a long-lived dust cloud with a radius of 9600km. Overall, we have a largely inconclusive result for Kepler-1520b given the data and our assumptions about the dust cloud variability. A function other than Raleigh may be needed, or else dust from below or above the star (impact parameter $> 1 R_{star}$) could be scattering more star light our direction than the extincted (scattered + absorbed) component.

\subsubsection{K2-22b\label{subsec:DatK2}}

\begin{figure}[t!]
\centering
\includegraphics[width=\columnwidth]{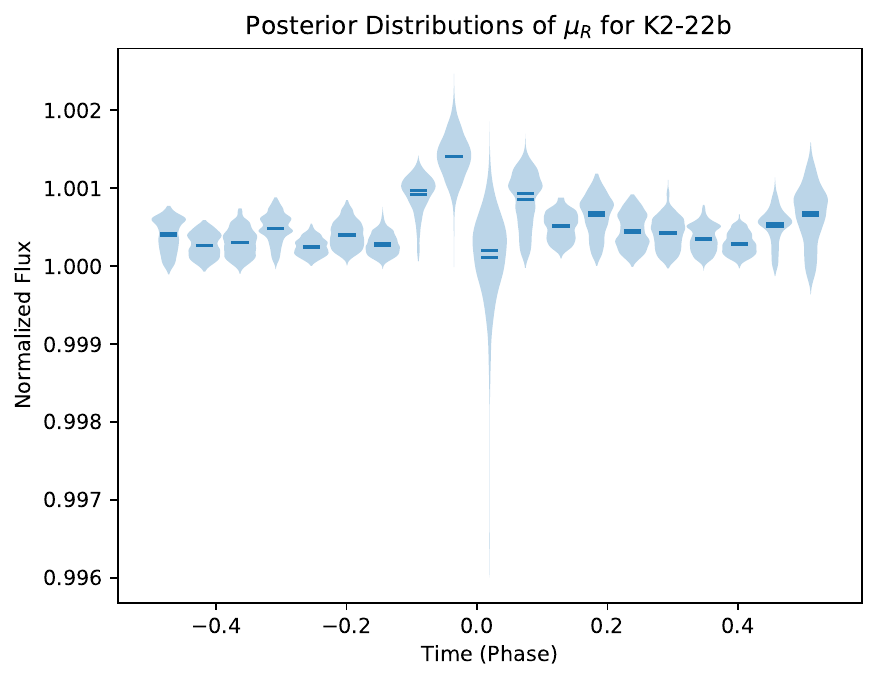}
\caption{Violin plot of $\mu_R$ for the posterior distributions of the joint function for K2-22b, where $\mu_R$ is the maximum of the Rayleigh component of the flux distribution. Each violin represents all the $\mu_R$ from a 30 minute segment of the orbit. \label{k2fig3}}
\end{figure}

\begin{figure}[t]
\centering
\includegraphics[width=\columnwidth]{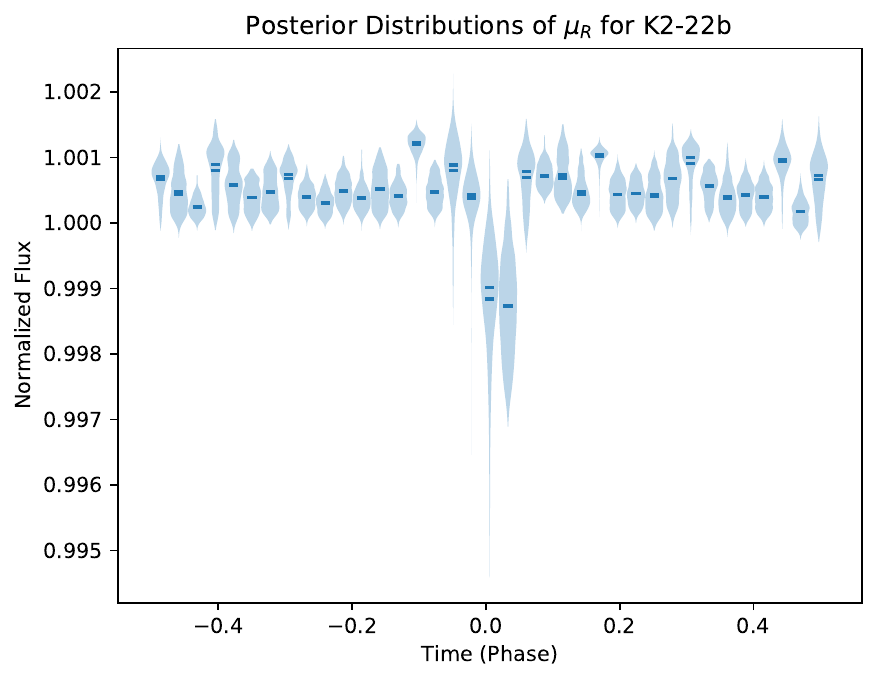}
\caption{Violin plots of $\mu_R$ for the posterior distributions of the joint function for K2-22b, where $\mu_R$ is the maximum of the Rayleigh component of the flux distribution. Each violin represents all the $\mu_R$ from a 15 minute segment of the orbit. \label{k2fig4}}
\end{figure}

Looking at the values of $\mu_R$ in Figures \ref{k2fig3} and \ref{k2fig4}, we can see that K2-22b has a dip that goes below the out-of-transit flux, indicating a planet transit. This appears more like a conventional planet transit of a non-disintegrating body. However, the increase in flux on either side of the dip in Figure \ref{k2fig3} indicates forward scattering.

Again, we need to calculate the transit duration of K2-22b for various impact parameters. We can use equation \ref{eq5},  stellar parameters from \citet{sanchis2015}, the upper limit of the exoplanet radius from \citet{2021AJ....162...57S} of $0.7 R_{\oplus}$, and the same values for the impact parameter of 0, 0.5, and 1 to find transit times of 53.5, 46.7 and 3.4 minutes respectively, or 0.098, 0.085, and 0.0047 of the orbit, again meaning that for all but a nearly grazing transit, our slices should time resolve the transit of K2-22b, as the transit should span one or two of the 30 minutes slices.

When looking at these graphs, two things become obvious. First, in Figure \ref{k2fig3}, there is a clear dip in flux below unity, meaning this is most likely an exoplanet transit. Second, when looking at Figures \ref{k2fig3} and \ref{k2fig4}, one can clearly see that there is an increase in flux after the transit, characteristic of a disintegrating exoplanet with a debris trail causing forward-scattering. Additionally, we see that there are two time slices in Figure \ref{k2fig4} that appear to contain the transit of K2-22b, indicating that the impact parameter is likely just below 0.5, meaning the transit is not grazing.

We want to put a constraint on the maximum size of K2-22b. First, when looking at Figure \ref{k2fig3}, we see the flux drops down and then goes  immediately back up, for an average change in flux of 0.097\%. Then, using Equation \ref{eq1} and $R_{Star} \approx 0.57R_\odot\ $ (\citet{sanchis2015}), we can place an upper limit of the radius of K2-22b at 12360km, which is slightly lower than the upper limit of the radius derived by \citet{sanchis2015}. However, this may be the result of the statistical fluctuation as well as bin location relative to the transit, so we also study the lightcurve with shorter duration time slices.

 If we study the posterior distribution with 15 minute slices in Figure \ref{k2fig4}, we can see that the transit takes place over two of the time slices. We will focus on the first slice of the transit which is more closely centered at Phase = 0, as it has a lower minimum value, indicating that it contains the main transit. Since the posterior distribution does not show much sign of an increase in flux before or after transit, we will instead use the average value of the out-of transit flux; that is, we will take the average of the fluxes for slices outside of the phase range $-0.2 < t < 0.2$, and compare the median of the transit slice to this value. Making this comparison, we find that the change in flux is 0.14\%. Using the same formula and value for the radius of the star, we find that the radius of the planet is approximately 14,850km ($2.4 R_{\oplus}$), which is nearly the same radius reported by \citet{sanchis2015}, but much larger than the radius of $\sim 0.7R_\oplus$ found in \citet{2021AJ....162...57S}. It should be noted that this method, from assumptions 1 \& 2 in Section \ref{joint_func} only removes the variable component of the dust. If a continuous low level dust cloud is present, it would likely appear similar to the transit of the underlying planet.
 
\subsection{Discussion}

\subsubsection{Kepler-1520b}

When looking at Figure \ref{kepviolin1}, we can see that during the transit, there is a difference in the shape between the observed flux and the joint function due to the elongated tails in the observed flux. This is due to the Rayleigh component having an abrupt truncation at one end, and a long, continually decreasing tail on the other. This is different than the observed fluxes, which appears to have segments where the counts stay constant as the flux decreases. This is due to a shortcoming of using the Rayleigh distribution and is an area for future research using more flexible functions to fit the astrophysical variation. However, the Rayleigh and Gaussian convolution still describes the general width and extent of the measured flux.

Additionally, in the violin plot on the right of Figure \ref{kepviolin1}, we should expect that the Rayleigh component of the joint function is a Delta function centered around flux = 1 during out-of-transit phases, but instead we find that the variability of the lightcurve extends into the assumed out of transit phases, from $Phase = -0.15$ to $Phase = 0.17$.
The most likely cause of the variation of this Rayleigh component is a small preceding dust trail and a longer dust tail behind Kepler-1520b, revealing additional variability in the system not detected in previous analysis of the lightcurves.

Figure \ref{kepfig3} shows the distribution of values for $\mu_R$, or the maximum of the Rayleigh component of our joint function. $\mu_R$ deviates significantly at the points from $Phase = -0.15$ to $Phase = 0.17$. We can see that everywhere that there is astrophysical variation, as indicated by a widening of the Rayleigh component from a delta function, the maximum of $\mu_R$ goes well above unity. This means that our assumptions that the transit is well-described by a Rayleigh and that the flux will decrease more than the increase from forward scattering during transit are flawed, as we should expect this value to go down during transit of the planet and the dust tails.

The reason we may be seeing an increase in $\mu_R$ during the in-transit phases is that the transit of Kepler-1520b may be a grazing transit (a transit with impact parameter $\gtrsim 1 R_{\odot}$, meaning that forward-scattering can have a greater impact on the amount of light collected than the obscuring of the light by the planet and the dust during transit. Additionally, our method of attempting to discern the size of the underlying planet using the average of many transits may be flawed, as forward-scattering by dust at different phases of the orbit may be increasing the measured flux, meaning that we may be greatly underestimating the amount of forward-scattering possible.

%
%



\subsubsection{Alternate Probability Distribution Functions}

In addition to a convolution of a Gaussian with a Rayleigh distribution, we also attempted to use PyMC3 to fit a truncated Gaussian - Exponential - Gamma convolution, a Gaussian - log-Normal convolution, and a Gaussian - logit-Normal convolution to the data. However, this analysis was unsuccessful, and models did not converge to a solution of the data regardless of the priors.

\section{Summary} \label{sec:summary}

In this paper, we have examined the disintegrating exoplanets Kepler-1520b and K2-22b in a new way in an attempt to place a better constraint on the maximum size of the exoplanets. We cleaned, de-trended, and phase-folded the data. From there, we were able to slice this phase-folded lightcurve into increments of time that were one time and two times the frequency of Kepler observations. We took these time slices and made histograms of the data, calculated Rayleigh and joint Gaussian-Rayleigh curve fits, where the Gaussian component of the joint curve fit was fixed to the out-of-transit distribution, and saved a plot for each slice. The fixed Gaussian component was calculated by compiling all of the data from $Phase \lesssim -0.2$ and $Phase \gtrsim 0.2$ and fitting a Gaussian to the resulting histogram

Lastly, we used PyMC3 and Bayesian inference to fit the lightcurve flux distribution and check the statistical significance of our results. We modeled the transiting exoplanets with a joint Gaussian and Rayleigh distribution, studying the posterior distributions of the joint function after fitting it to our observed data. We detect a transit-like dip in the dust-removed lightcurve of K2-22b of 0.14\% that corresponds to a radius of $2.4R_{\oplus}$, but were unable to detect a robust transit dip for Kepler-1520b. Since the maximum of the Rayleigh distribution is greater than 1.0 during the in-transit phase for Kepler-1520b, our results are dominated by forward scattering of dust, meaning that the transits may be only be grazing transits, or that forward scattering from dust in other parts of the orbit obscures the transit from the underlying planet. However, the increase of the widths of the Rayleigh functions prior to the transit of K2-22b is a new result obtained via our method, indicating astrophysical variation, or a dust trail, preceding the transit. Therefore, continued research may be warranted with a new distribution modeling the astrophysical variation.

We looked at the maximum time series of each exoplanet in an attempt to put a new constraint on the maximum size of the planets for two different time slice sizes, and found hints of transit features for the 15 minute time slices. There is a statistically significant dip in flux for K2-22b with an upper limit of $2.4R_{\oplus}$, confirming previous upper limits on the radius. For Kepler-1520b, our assumption that the dust causes a negative flux dip following a Raleigh distribution is incorrect, hinting at either a grazing geometry of the orbit or that a more complex astrophysical variation model is needed.

\acknowledgments

Funding for ES is supported by the NASA Goddard Spaceflight center.
We respectfully acknowledge the University of Arizona is on the land and territories of Indigenous peoples. Today, Arizona is home to 22 federally recognized tribes, with Tucson being home to the O'odham and the Yaqui. Committed to diversity and inclusion, the University strives to build sustainable relationships with sovereign Native Nations and Indigenous communities through education offerings, partnerships, and community service.

The data underlying this article are available in Kepler Mission Data Resources in the Exoplanet Archive, at  \url{https://exoplanetarchive.ipac.caltech.edu/docs/KeplerMission.html}, and can be accessed with Kepler IDs: 12557548, and 201637175.

\software{\texttt{astropy \citep{astropy}, numpy \citep{numpy}, lightkcurve \citep{Lightkurve}, pymc3 \citep{pymc3}, and scipy \citep{2020SciPy-NMeth}}}


\appendix
All code used to compute our values and generate our plots can be found at \citet{Github}.

\bibliography{sample63}{}
\bibliographystyle{aasjournal}

\allauthors

\listofchanges

\end{document}